\documentclass[pra,aps,showpacs,groupedaddress,superscriptaddress,onecolumn,longbibliography,notitlepage]{revtex4-1}

\usepackage[utf8x]{inputenc}
\usepackage{color}
\usepackage{bbm} 

\usepackage{amsfonts,amsmath,amssymb,stmaryrd}

\usepackage{graphicx}
\usepackage{subfigure}  
\usepackage{bbm} 
\usepackage{hyperref}
\usepackage[pdftex]{epsfig}
\usepackage{mathrsfs}
\usepackage{verbatim}
\usepackage{centernot}
\usepackage{ulem}
\usepackage[extra]{tipa}

\renewcommand{\l}{\left(}
\renewcommand{\r}{\right)}

\newcommand{\hc}{\text{h.c.}}

\newcommand{\MF}{{\rm MF}}

\newcommand{\ket}[1]{|#1\rangle}

\renewcommand{\H}{\hat{\mathcal{H}}}

\renewcommand{\a}{\hat{a}}

\newcommand{\ad}{\hat{a}^\dagger}

\newcommand{\Omt}{\tilde{\Omega}}
\newcommand{\G}{\hat{\Gamma}}

\newcommand{\Gt}{\hat{\tilde{\Gamma}}}

\newcommand{\Gtt}{\hat{\text{\doubletilde{$\Gamma$}}}}

\newcommand{\f}{\text{f}}

\newcommand{\ph}{\text{ph}}

\newcommand{\IB}{\text{IB}}
\newcommand{\p}{{\rm p}}
\newcommand{\U}{\hat{U}}
\newcommand{\Ud}{\hat{U}^\dagger}
\newcommand{\F}{\hat{F}}
\renewcommand{\s}{{\rm s}}

\usepackage{array}

\usepackage{cancel,ifthen}
\newcommand{\cmnt}[2][NoInPuT]{\ifthenelse{\equal{#1}{NoInPuT}}{}{{\color{red}\sout{#1}}} {\color{blue} #2}}

\usepackage{bm}	
\renewcommand{\vec}[1]{\bm{#1}}

\begin{document}
\normalem	

\title{An all-coupling theory for the Fr\"ohlich polaron}

\author{Grusdt, F.}
\email[Corresponding author email address: ]{grusdt@physik.uni-kl.de}
\affiliation{Department of Physics, Harvard University, Cambridge, Massachusetts 02138, USA}
\affiliation{Department of Physics and Research Center OPTIMAS, University of Kaiserslautern, Germany}

\begin{abstract}
The Fr\"ohlich model describes the interaction of a mobile impurity with a surrounding bath of phonons which leads to the formation of a quasiparticle, the polaron. In this article an efficient renormalization group approach is presented which provides a description of Fr\"ohlich polarons in all regimes ranging from weak- to strong coupling. We apply the method to the Bose polaron problem of an ultracold impurity atom interacting with a background gas that is Bose-condensed. The extended renormalization group approach introduced here is capable to predict ground state properties for arbitrarily small impurity masses. This allows us to obtain the full phase diagram of the corresponding Bogoliubov-Fr\"ohlich Hamiltonian, characterized by two dimensionless coupling constants. Our method is benchmarked by comparison of the ground state energy to recent diagrammatic quantum Monte Carlo calculations.
\end{abstract}



\date{\today}

\maketitle

\section{Introduction}

When an electron moves through a metal, it interacts with the phonons of the host lattice. Such electron-phonon interactions were first described by Fr\"ohlich who introduced a model where the electron emits and reabsorbs phonons \cite{Froehlich1954}. When a single electron is considered as a mobile impurity in the crystal, as noted by Landau and Pekar \cite{Landau1946,Landau1948}, it becomes dressed by a cloud of phonons which it carries along. The resulting quasiparticle is called a polaron, and compared to the bare electron it has an enhanced mass. Such polaronic mass enhancement is not limited to electrons in solids but is now understood as a generic effect that can be observed in much broader classes of systems, including magnets \cite{Auerbach1991,Berciu2009} or ultracold atoms \cite{Mathey2004,Bruderer2007,Catani2012,Fukuhara2013,Grusdt2015Varenna}.

Polarons are mostly considered in one out of two physical regimes. For weak couplings the mobile impurity retains is character and its mass is slightly increased because of the dressing with virtual phonons. For strong couplings, on the other hand, the impurity becomes self-trapped inside a sizable potential created by the phonons making up the polaron cloud. Such strong-coupling polarons are extremely heavy and loose their character as mobile impurities almost entirely. The most interesting physics takes place at intermediate couplings, however, where the impurity acts as an exchange particle mediating interactions between phonons. As a result the phonons become correlated in this regime and a non-trivial polaronic state is expected \cite{Tulub1962,Grusdt2015RG,Shchadilova2014}. In this article we introduce a theoretical method, valid for arbitrary couplings, to describe such correlations quantitatively.

In both regimes of weak and strong coupling efficient theories exist for describing polarons, see Refs.\cite{Devreese2013,Grusdt2015Varenna} for reviews. At strong couplings Landau and Pekar \cite{Landau1946,Landau1948} used the adiabatic principle and assumed that phonons can follow the impurity instantly. At weak couplings, on the other hand, simple perturbation theory or the more advanced mean-field (MF) approach developed by Lee, Low and Pines \cite{Lee1953} can be used. Although the theory of polarons is now more than eighty years old, the regime of intermediate couplings is still poorly understood and efficient methods for describing polarons at such couplings are lacking. The most accurate method is the diagrammatic Monte Carlo approach \cite{Prokofev1998} but it only provides limited physical insights. The most successful semi-analytical all-coupling theory in cases where optical phonons are involved \cite{Peeters1985Feynman} was developed by Feynman \cite{Feynman1955}. He introduced a variational theory, formulated in terms of path integrals, allowing to extrapolate between the weak and strong coupling limits where the Lee-Low-Pines and Landau-Pekar theories are asymptotically recovered, respectively. Due to its restriction to a set of only two variational parameters it is unable to capture all the quantum correlations between phonons which are present at intermediate couplings however. This was demonstrated by recent diagrammatic Monte Carlo calculations of the energy of an impurity atom inside an atomic Bose-Einstein condensate (BEC), where large discrepancies were found to predictions by Feynman's variational method \cite{Vlietinck2015}.

\begin{figure}[t]
\centering
\epsfig{file=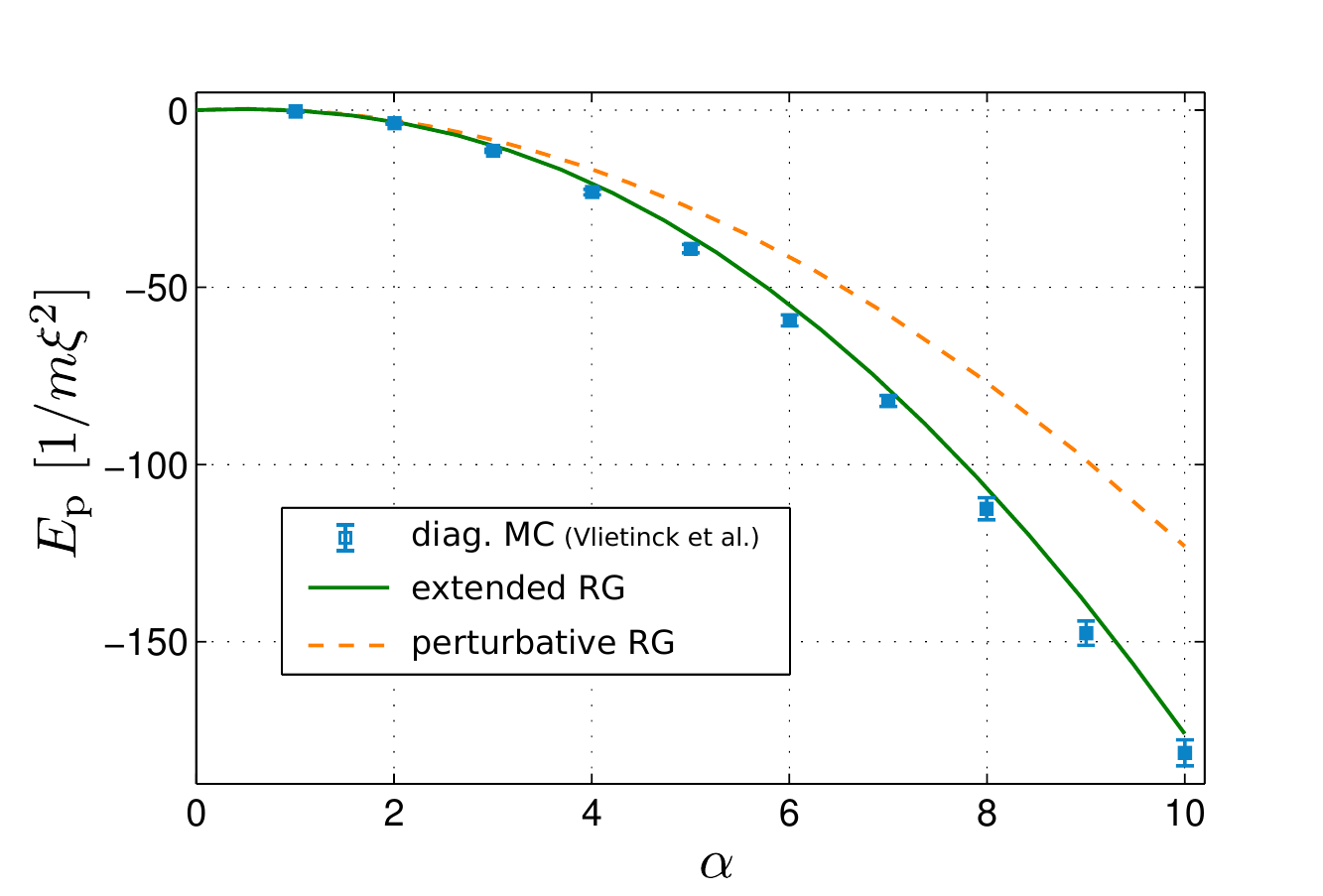, width=0.5\textwidth}
\caption{To investigate polarons at arbitrary couplings, we extend the perturbative RG approach introduced in Ref.\cite{Grusdt2015RG}. Here we compare polaron ground state energies $E_{\rm p}$ of an impurity atom in a three-dimensional BEC, calculated using different techniques. The extended RG scheme reproduces accurately the energy predicted by numerical diagrammatic Monte Carlo calculations by Vlietinck et al. \cite{Vlietinck2015} for arbitrary couplings. The simpler perturbative RG approach \cite{Grusdt2015RG} agrees with these results only for moderate values of $\alpha \lesssim 4$. Parameters are $M/m = 0.26316$, $\Lambda_0=3000/\xi$ and $P=0$ for all curves.}
\label{fig:extendedRGenergies}
\end{figure}

Recently a new theoretical method was introduced, based on the use of renormalization group (RG) techniques, to include phonon-phonon correlations in the description of the polaron cloud \cite{Grusdt2015RG,Grusdt2015Varenna}. It was employed to explain the discrepancies between Feynman path integral calculations and numerically exact diagrammatic Monte Carlo results \cite{Vlietinck2015} by predicting the existence of a peculiar logarithmic divergence of the BEC-polaron energy with the ultra-violet (UV) momentum cut-off. It was shown that Feynman's method as well as the celebrated strong-coupling theory of Landau and Pekar are unable to capture the effects of quantum fluctuations in the UV which lead to the log-divergence \cite{Grusdt2015Varenna}. The RG technique developed in Ref.\cite{Grusdt2015RG} not only yields analytical insights to the polaron problem but it also provides an accurate quantitative description of the ground state from weak to intermediate couplings. 

We have identified two main short-comings of the perturbative RG described in Ref.\cite{Grusdt2015RG}, which will be overcome by the extended RG approach introduced in this article. First, for strong couplings $\alpha \gg 1$ (where $\alpha$ is the dimensionless coupling constant) the ground state energies obtained from the perturbative RG compare poorly with diagrammatic Monte Carlo results, see FIG.\ref{fig:extendedRGenergies}. Second, in the regime of light impurities at strong couplings, where quantum fluctuations are dominant, the perturbative RG predicts an unphysical divergence of the effective polaron mass, similar to the prediction of a simple perturbation theory in the impurity-phonon coupling. In this article we show that both these problems originate from the insufficiency of using only perturbation theory in the derivation of the RG flow. 

We extended the RG by going beyond perturbation theory \emph{in every RG step}, each time effectively summing up an infinite number of diagrams. As a result we obtain an efficient semi-analytical theory for the polaron which provides an accurate quantitative description of the ground state for arbitrary coupling strengths, see FIG.\ref{fig:extendedRGenergies}. The key insight is to expand the renormalized Hamiltonian around its new MF ground state after every RG step. This approach is not restricted to Fr\"ohlich Hamiltonians, but may be applied to much broader classes of systems to obtain a quantitative description of correlated ground states from the renormalization group approach in general.

The article is organized as follows. In section \ref{sec:Model} we introduce the Bogoliubov-Fr\"ohlich model describing polarons in a weakly interacting BEC which we will use to construct the extended RG approach. In Sec. \ref{sec:MethodOverview} we provide an overview of our method and introduce the coupling constants flowing in the RG protocol. In Sec.\ref{sec:Results} we discuss our results for impurity atoms inside a BEC and compare them to quantum Monte Carlo calculations and results obtained using Feynman's variational method. We calculate the phase diagram of the Bogoliubov-Fr\"ohlich model for all values of the two dimensionless coupling constants in two spatial dimensions. Its qualitative shape in the regime where the impurity is much lighter than host bosons is explained in Sec.\ref{sec:SmallImptMass}. In Sec.\ref{sec:extRG} we present the detailed derivation of the extended RG method. We close with a summary and by giving an outlook in Sec.\ref{sec:SummaryOutlook}.

\section{Model}
\label{sec:Model}

Traditionally the polaron problem was investigated in the context of solid state physics, however recently impurity atoms immersed in ultracold quantum gases have become an increasingly important model system for studying polaronic effects \cite{Mathey2004,Bruderer2007,Catani2012,Fukuhara2013,Grusdt2015Varenna}. These ultra clean systems are ideally suited for an investigation of polaron physics because of the powerful experimental tools available and the tunability of many model parameters over a wide range. The impurity phonon interactions can be controlled using Feshbach resonances \cite{Chin2010} and the effective impurity mass can be tuned almost at will by coupling to light fields \cite{Grusdt2015DSPP}. This motivates us to study a model relevant to ultracold atoms in the following, although our method can be applied to more general classes of systems. 

A single impurity atom interacting with a surrounding Bose-Einstein condensate (BEC) can be described by the Bogoliubov-Fr\"ohlich Hamiltonian
\begin{equation}
\H =  \frac{\hat{\vec{p}}^2}{2 M}  + \int^{\Lambda_0} d^d \vec{k} ~ \Biggl[  V_{k} e^{i \vec{k}\cdot \hat{\vec{r}}} \l \a_{\vec{k}} + \ad_{-\vec{k}} \r  + \omega_{k} \ad_{\vec{k}} \a_{\vec{k}}  \Biggr]
\label{eq:HFroh}
\end{equation}
in $d$ spatial dimensions, under conditions where the condensate depletion in the vicinity of the impurity remains small \cite{Bruderer2007,Tempere2009,Grusdt2015Varenna}. Here $\hat{\vec{p}}$ ($\hat{\vec{r}}$) denote the impurity momentum (position) operators respectively, and $\a_{\vec{k}}$ annihilates a Bogoliubov excitation of the BEC. The mass of the impurity is $M$ and $\Lambda_0$ is a UV momentum cut-off required for regularization \cite{Grusdt2015Varenna}. The scattering amplitude is defined by
\begin{equation}
V_k = \sqrt{\alpha} \frac{c \sqrt{\xi}}{2 \pi \sqrt{2}} \l 1 + \frac{m_{\rm B}}{M} \r  \l \frac{k^2 \xi^2}{2 + k^2 \xi^2} \r^{1/4} 
\end{equation}
where $\alpha = a_{\rm IB}^2 / (a_{\rm BB} \xi)$ is the 3d dimensionless coupling constant \cite{Tempere2009}, $a_{\IB}$ ($a_{\rm BB}$) is the impurity-boson (boson-boson) scattering length, $\xi$ and $c$ denote the healing length and the speed of sound in the BEC and $m_{\rm B} = 1 /( \sqrt{2} c \xi )$ is the mass of bosons in the BEC. The Bogoliubov dispersion is given by $\omega_k=c k \sqrt{1 + k^2 \xi^2/2}$. The dependence of $V_k$ and $\omega_k$ on the momentum $k$ is specific for the BEC polaron, but our theoretical analysis below applies to any Hamiltonian of the type in Eq.\eqref{eq:HFroh}, in any spatial dimension $d$. 

The healing length $\xi$ defines the natural length scale of the Bogoliubov-Fr\"ohlich Hamiltonian and $c/\xi$ the characteristic energy scale. This leaves only two independent dimensionless coupling constants $\alpha$ and $m_{\rm B}/M$ which fully characterize the model for a given UV cut-off $\Lambda_0$.

\section{Method -- an overview}
\label{sec:MethodOverview}

In this section we provide a brief overview of our theoretical treatment of the Bose polaron problem. We start from the general Fr\"ohlich Hamiltonian \eqref{eq:HFroh} and apply the unitary Lee-Low-Pines transformation \cite{Lee1953}
\begin{equation}
\hat{U}_\text{LLP} = e^{i \hat{S}} ,\qquad \hat{S} = \hat{\vec{r}} \cdot \hat{\vec{P}}_{\rm ph},
\label{eq:LLPdef}
\end{equation}
where the total phonon momentum is given by $\hat{\vec{P}}_{\rm ph} = \int d^dk ~ \vec{k}  ~\ad_{\vec{k}} \a_{\vec{k}}$. We thus switch into the polaron frame where the impurity is localized in the origin. In the new basis the Hamiltonian commutes with $\hat{\vec{p}} = \vec{P}$, which becomes a $\mathbb{C}$-number and corresponds to the total momentum $\vec{P}$ of the polaron. It is conserved because of the translational invariance of the system. For more details, see e.g. Ref. \cite{Grusdt2015Varenna}.

Then, as in the perturbative RG \cite{Grusdt2015RG}, we expand around the MF saddle-point solution of the resulting Hamiltonian $\H_{\vec{P}} = \hat{U}_{\rm LLP}^\dagger \H \hat{U}_{\rm LLP}$. The MF solution corresponds to a wavefunction where the phonons are in coherent states $\ket{\alpha_{\vec{k}}^\MF}$. The MF amplitude is given by \cite{Shashi2014RF} $\alpha_{\vec{k}}^\MF = - V_k / \Omega_{\vec{k}}^\MF$ where the phonon dispersion in the new frame is
\begin{equation}
\Omega^\MF_{\vec{k}} = \omega_k + \frac{k^2}{2 M} - \frac{1}{M } \vec{k} \cdot \l \vec{P} - \vec{P}_\ph^\MF \r, \qquad {\rm and} \qquad \vec{P}_\ph^\MF = \int d^d\vec{k} ~ \vec{k} |\alpha^\MF_{\vec{k}}|^2
\label{eq:OmegakDef}
\end{equation}
denotes the MF phonon momentum. We change into the frame of quantum fluctuations by applying the unitary MF shift $\hat{U}_{\rm MF} = \exp \l \int d^d \vec{k} ~ \alpha^{\rm MF}_{\vec{k}} ~ \ad_{\vec{k}} - \hc \r$ and end up with the following equivalent representation of the Fr\"ohlich Hamiltonian \eqref{eq:HFroh}
\begin{equation}
\tilde{\mathcal{H}} = \hat{U}^\dagger_\MF  \hat{U}^\dagger_{\rm LLP} \H \hat{U}_{\rm LLP} \hat{U}_{\rm MF} = E_0^\MF + \int^{\Lambda_0} d^d \vec{k} ~ \ad_{\vec{k}} \a_{\vec{k}} \Omega^\MF_{\vec{k}}  + \int^{\Lambda_0} d^d \vec{k} ~ d^d \vec{k}' ~  \frac{\vec{k} \cdot \vec{k}'}{2 M}  ~ : \G_{\vec{k}} \G_{\vec{k}'} :.
\label{eq:HquantFlucDef}
\end{equation}
Here $E_0^\MF$ denotes the variational MF ground state energy and we defined the operators $\G_{\vec{k}} := \alpha^\MF_{\vec{k}} ( \a_{\vec{k}} + \ad_{\vec{k}} ) + \ad_{\vec{k}} \a_{\vec{k}}$. With $:...:$ we denote normal-ordering of phonon operators, e.g. $:\a_{\vec{k}} \ad_{\vec{k}'}: ~ = \ad_{\vec{k}'} \a_{\vec{k}}$ and $:\ad_{\vec{k}} \a_{\vec{k}'}:~  = \ad_{\vec{k}} \a_{\vec{k}'}$.

\subsection{Extended RG method}
\label{eq:overviewExtRG}
The Hamiltonian \eqref{eq:HquantFlucDef} is the starting point for all further analysis. In Refs.\cite{Grusdt2015RG,Grusdt2015Varenna} a RG approach based on phonon wavefunctions, related to Wegner's method \cite{Wegner1994}, was introduced to describe the ground state of this Hamiltonian. The key idea is to distinguish between fast phonons with momenta $\vec{k}$ form a thin momentum shell $\Lambda -\delta \Lambda < |\vec{k}| < \Lambda$, and slow phonons with momenta $|\vec{p}| < \Lambda - \delta \Lambda$. The typical frequency $\Omega_{\vec{k}}$ of fast phonons sets the largest energy scale, and the goal is to decouple fast and slow degrees of freedom perturbatively in $\Omega_{\vec{k}}^{-1}$. This generates an RG flow of the Hamiltonian.

In practice the decoupling can be performed by applying a series of unitary transformations $\hat{U}_\Lambda$ at each momentum shell. This was done already in the perturbative RG of Ref. \cite{Grusdt2015RG}, and it was found that additional terms are generated in the Hamiltonian \eqref{eq:HquantFlucDef} which are of the type $W_{\vec{k}} ( \a_{\vec{k}} + \ad_{\vec{k}} )$ with some function $W_{\vec{k}}$. As a consequence the interactions between phonons, described by the last term in Eq.\eqref{eq:HquantFlucDef}, no longer involve small fluctuations around the MF saddle point solution of the new Hamiltonian. Instead they involve phonons $\a_{\vec{k}}$ which are subject to the displacement $W_{\vec{k}}$ growing during the RG flow. 

To ensure that the perturbative decoupling of fast and slow phonons involves the smallest possible fluctuations, we apply a second unitary transformation
\begin{equation}
\hat{V}_\MF(\Lambda) = \exp \l \int^\Lambda d^d \vec{p} ~  \delta \tilde{\alpha}_{\vec{p}} ~  \ad_{\vec{p}} -  \hc \r
\label{eq:UMF-RG}
\end{equation}
after every RG step, by which we change into the basis of fluctuations around the saddle point. This is done by properly choosing $\delta \tilde{\alpha}_{\vec{p}}$. Hence we displace the slow phonons by an amount $\delta \tilde{\alpha}_{\vec{p}}$ which effectively renormalizes the MF amplitude in very RG step. This is how the extended RG differs from its purely perturbative cousin. 

\subsubsection{Universal Hamiltonian}
As a result of the extended RG, the universal Hamiltonian takes a simpler form than in the perturbative case \cite{Grusdt2015RG}. Below in Sec.\ref{sec:extRG} we will derive RG flow equations for the coupling constants of the following universal Hamiltonian at UV cut-off $\Lambda$,
\begin{equation}
\tilde{\mathcal{H}}_P(\Lambda) = E_0 (\Lambda) + \int^\Lambda d^d \vec{k} ~ \ad_{\vec{k}} \a_{\vec{k}} \Omt_{\vec{k}}  
+ \int^\Lambda d^d \vec{k} ~ d^d \vec{k}' ~ \frac{1}{2}  k_\mu \mathcal{M}_{\mu \nu}^{-1} k_\nu'  ~ : \Gt_{\vec{k}} \Gt_{\vec{k}'} :.
\label{eq:HquantFluc2}
\end{equation}
Here $\mu, \nu=x,y,...$ denote spatial coordinates which are summed over using Einstein convention. We defined the operators $\Gt_{\vec{k}} := \tilde{\alpha}_{\vec{k}} ( \a_{\vec{k}} + \ad_{\vec{k}} ) + \ad_{\vec{k}} \a_{\vec{k}}$, where the MF amplitude $\tilde{\alpha}_{\vec{k}}$ is flowing in the RG. The solution $\tilde{\alpha}_{\vec{k}} = - V_k / \Omt_{\vec{k}}$ is reminiscent of the MF expression, but the renormalized frequency is given by 
\begin{equation}
\Omt_{\vec{k}} = \omega_k + \frac{1}{2} k_\mu \mathcal{M}_{\mu \nu}^{-1} k_\nu + k_\mu \mathcal{M}^{-1}_{\mu \nu} \kappa_\nu.
\end{equation}
The two coupling constants of the RG are given by the momentum $\kappa_\mu$ and the imupurity mass tensor $\mathcal{M}_{\mu \nu}$. Comparison to Eq.\eqref{eq:HquantFlucDef} shows that the initial conditions for the RG are 
\begin{equation}
\kappa_\mu(\Lambda_0) = \delta_{\mu x} \l P_{\rm ph}^{\rm MF} - P \r, \qquad \mathcal{M}_{\mu \nu}(\Lambda_0) = \delta_{\mu \nu} M^{-1}, \qquad E_0(\Lambda_0) = E_0 |_{\rm MF}.
\end{equation}
Here we found it convenient to assume that the total system momentum $\vec{P} = P \vec{e}_x$ points along the $x$-direction. For symmetry reasons also the MF momentum has to point in this direction, $\vec{P}_\ph^\MF = \vec{e}_x P_\ph^\MF$.

\subsubsection{RG flow equations}
In Sec.\ref{sec:extRG} we will derive the following RG flow equations,
\begin{equation}
\frac{\partial \mathcal{M}_{\mu \nu}^{-1} }{\partial \Lambda} = 2 \mathcal{M}_{\mu \lambda}^{-1}  \int_\f d^{d-1} \vec{k} ~ \frac{V_k^2}{\Omt^3_{\vec{k}}} k_\lambda k_\sigma  ~\mathcal{M}_{\sigma \nu}^{-1}
\label{eq:gsFlowM} 
\end{equation}
for the mass (where $\int_\f d^{d-1} \vec{k}$ denotes the integral over the $d-1$-dimensional momentum shell with radius $\Lambda$), and for the momentum
\begin{equation}
\frac{\partial \kappa_x }{\partial \Lambda} = - \frac{\partial \mathcal{M}_{\mu \nu}^{-1} }{\partial \Lambda} \frac{I^{(3)}_{\mu \nu}}{1 + 2 \mathcal{M}_\parallel^{-1} I^{(2)}} - \frac{\partial \mathcal{M}_\parallel^{-1}}{\partial \Lambda} \mathcal{M}_\parallel \kappa_x.
\label{eq:MFRGflowkappaX}
\end{equation}
Here $\mathcal{M}_\parallel = \mathcal{M}_{xx}$ denotes the component of the tensorial mass along the direction of the total polaron momentum $\vec{P} = P \vec{e}_x$. Furthermore we defined the following integrals,
\begin{equation}
I^{(2)}(\Lambda) = \int^\Lambda d^d \vec{k} ~ k_x^2 \frac{V_k^2}{\Omt^3_{\vec{k}}}, \qquad \qquad I^{(3)}_{\mu \nu} (\Lambda) =  \int^\Lambda d^d \vec{k} ~ k_x k_\mu k_\nu \frac{V_k^2}{\Omt^3_{\vec{k}}}.
\end{equation}

For the ground state energy we derive the following RG flow equation,
\begin{equation}
\frac{\partial E_0}{\partial \Lambda} = \frac{1}{2} \frac{\partial \mathcal{M}_{\mu \nu}^{-1}}{\partial \Lambda} \int_\s d^d \vec{p} ~ p_\mu p_\nu \l \tilde{\alpha}_{\vec{p}} \r^2.
\label{eq:MFRGflowEnergy}
\end{equation}
This expression was used to calculate the polaron energies shown in FIG.\ref{fig:extendedRGenergies}.

\subsection{Effective polaron mass}
Next we explain how to calculate the effective polaron mass $M_\p$ form the extended RG approach. It is defined by expanding the ground state energy $E_0(P)$ with respect to the polaron moment, $E_0(P) = E_0(0) + P^2 / 2 M_\p + \mathcal{O}(P^4)$. In Ref.\cite{Grusdt2015RG} we showed using Ehrenfest's theorem that the polaron mass can be related to the total phonon momentum in the ground state $P_\ph = \int d^d \vec{k} ~ k_x \langle \ad_{\vec{k}} \a_{\vec{k}} \rangle$,
\begin{equation}
M_{\rm p} = M  \lim_{P \to 0} \l 1 - P_{\rm ph} / P \r^{-1}.
\end{equation}

The total polaron momentum $P$ can be partitioned into a part $P_{\rm I}$ carried by the impurity and a part $P_{\rm ph}$ carried by the phonon cloud, $P = P_{\rm I} + P_{\rm ph}$. Intuitively one expects that both contributions are positive, $P_{\rm I}, P_{\rm ph} > 0$, which leads to a positive polaron mass $M_{\rm p} > 0$. Indeed, within the fully self-consistent MF theory one can proof that this is always the case, see e.g. Ref.\cite{Shashi2014RF}. However in the perturbative RG of Ref. \cite{Grusdt2015RG} the phonon momentum $P_{\rm ph}$ is calculated non self-consistently. We found that small systematic errors $\delta P_{\rm ph}$ lead to slightly negative impurity momenta $P_{\rm I} = P - P_{\rm ph} < 0$ for light impurities in the strong-coupling regime. These, in turn, may have the dramatic effect of predicting a large but negative mass for what should be a very heavy polaron but with a positive mass. In the extended RG approach, on the other hand, the phonon momentum is treated self-consistently because in every RG step the MF amplitude is readjusted, see Eq. \eqref{eq:UMF-RG}.

The RG flow equation for the phonon momentum can be derived using the techniques introduced in Ref.\cite{Grusdt2015Varenna}. Using the self-consistent RG scheme we obtain
\begin{equation}
\frac{\partial P_\ph}{\partial \Lambda} = - \frac{\partial \mathcal{M}_{\mu \nu}^{-1}}{\partial \Lambda} \frac{I_{\mu \nu}^{(3)}}{1 + 2 \mathcal{M}_{\parallel}^{-1} I^{(2)}} \frac{M}{\mathcal{M}_{\parallel}}, \qquad \qquad P_\ph(\Lambda_0) = P_\ph^\MF.
\end{equation}
This differential equation can readily be solved by noting the close relation to the RG flow equation \eqref{eq:MFRGflowkappaX}. The solution is $P_\ph(\Lambda) = \kappa_x(\Lambda)  M / \mathcal{M}_\parallel (\Lambda)+ P$. From the last result we can now derive for the polaron mass,
\begin{equation}
M_\p = - \lim_{P \to 0} ~ \frac{P}{\kappa_x} \mathcal{M}_\parallel > 0.
\end{equation}
Notably, from this expression it follows analytically that the polaron mass is always positive. To see this, note that both $P$ and $\mathcal{M}_\parallel$ are always positive. $\kappa_x$, on the other hand, starts out negative. It stays negative during the entire RG flow, because for $\kappa_x=0$ we obtain $I^{(3)}_{\mu \nu}=0$ and thus there is no RG flow which could make $\kappa_x$ become positive, see Eq.\eqref{eq:MFRGflowkappaX}.

\section{Results}
\label{sec:Results}

Now we apply the extended RG method to the Bose polaron problem of an ultracold impurity inside a BEC. To benchmark our method we compare ground state energies obtained by the RG to recent diagrammatic quantum Monte Carlo calculations \cite{Vlietinck2015}. The results are shown in FIG.\ref{fig:extendedRGenergies}. The extended RG is in excellent agreement with numerically exact Monte Carlo predictions, where deviations are within the errorbars. This is the case for all values of the coupling constant ranging well into the strong coupling regime. For comparison, Feynman's variational approach predicts a transition from weak to strong coupling around $\alpha \approx 3$ for the parameters used in the Figure, see Ref.\cite{Tempere2009}. Around this point we also start to observe deviations of the perturbative RG scheme \cite{Grusdt2015RG} from the extended RG.

\begin{figure}[t]
\centering
\epsfig{file=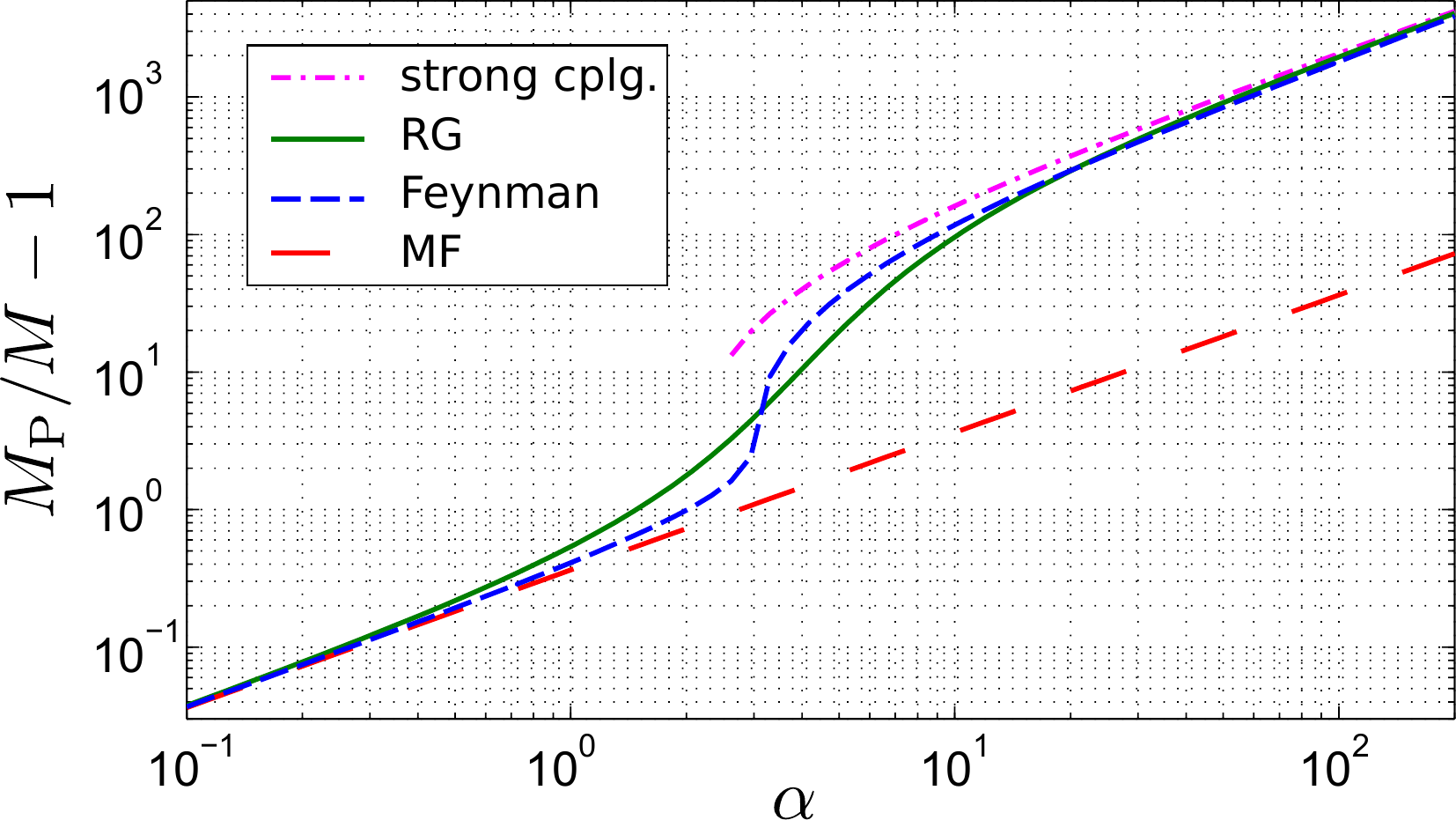, width=0.5\textwidth}
\caption{We show the dependence of the effective polaron mass $M_{\rm p}$ on the coupling constant $\alpha$ in $d=3$ dimensions. The extended RG reproduces correctly the asymptotically exact results of MF (weak-coupling) and Landau-Pekar theory (strong coupling). Parameters are $\Lambda_0=200/\xi$, $P=0.01 M c$, $M/m=0.26$ as in Refs.\cite{Tempere2009,Grusdt2015RG}.}
\label{fig:extendedRGpolaronMass}
\end{figure}

In FIG.\ref{fig:extendedRGpolaronMass} we show the effective polaron mass as a function of $\alpha$. We observe that the extended RG scheme reproduces correctly the polaron mass at large couplings, where Landau and Pekar's strong-coupling approach becomes accurate \cite{Landau1946,Landau1948,Casteels2011}. This is remarkable because the RG is constructed as an expansion around the weak coupling solution, where correspondingly the MF prediction is asymptotically approached. The perturbative RG \cite{Grusdt2015RG}, on the other hand, failed to reproduce the polaron mass correctly at large couplings. Unlike in the extended RG the flow of the phonon momentum was treated perturbatively, leading to a divergence of the polaron mass for light impurities. For heavy impurities the perturbative RG did however capture correctly the qualitative dependence of $M_{\rm p}$ on $\alpha$, see Ref.\cite{Grusdt2015RG}. 

\begin{figure}[b]
\centering
\epsfig{file=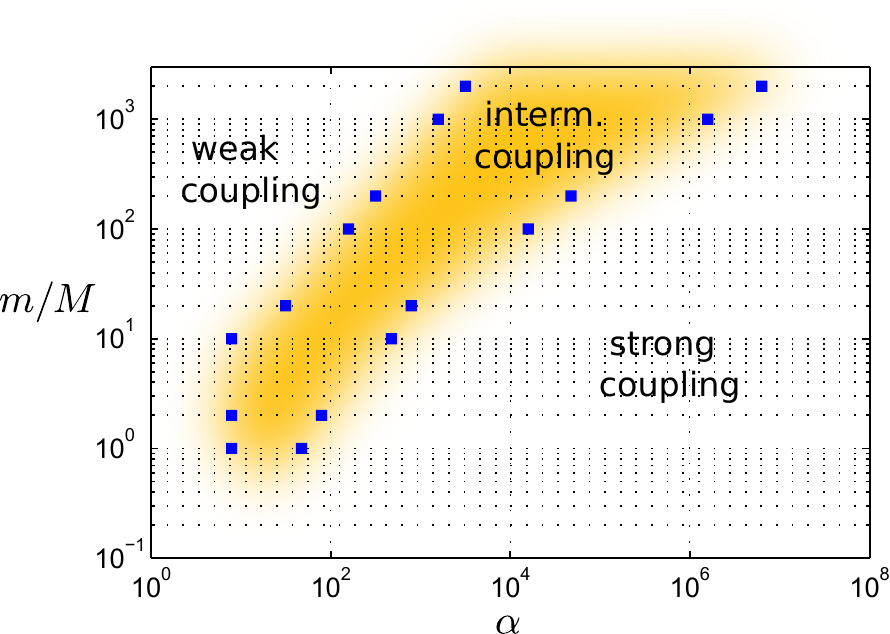, width=0.5\textwidth}
\caption{Full phase diagram of Bogoliubov-Fr\"ohlich model in two dimensions. This result was also presented in Ref.\cite{Grusdt2015DSPP}.}
\label{fig:setup}
\end{figure}

Unlike Feynman's variational method, which predicts a sharp transition from weak to strong couplings (note the double-logarithmic scale in FIG.\ref{fig:extendedRGpolaronMass}), we find a smooth cross-over with an extended regime of intermediate couplings. The same behavior was predicted by the perturbative RG in Ref.\cite{Grusdt2015RG}. Both for weak and strong couplings the polaron mass depends linearly on $\alpha$. Loosely we define the intermediate coupling regime by regions where the dependence of the polaron mass on $\alpha$ is non-linear.

In this way we can obtain the phase diagram of the Fr\"ohlich Hamiltonian. In FIG. \ref{fig:setup} we present our results for the two-dimensional Bogoliubov Fr\"ohlich model relevant e.g. for photonic polarons inside a quasi-two dimensional BEC \cite{Grusdt2015DSPP}. The qualitative shape of the phase diagram is the same in three dimensions. For sufficiently light impurities we find a large regime of intermediate couplings. It is located at increasingly larger $\alpha$ as the impurity mass $M$ is decreased. Indeed, we will show in the next section by an exact calculation that MF theory becomes exact in the limit $M \to 0$ and for any fixed $\alpha$. For heavy impurities the impurity becomes localized and weak and strong coupling theories are equivalent.

In two dimensions the coupling constant $\alpha$ defines the scattering amplitude as $V_k = \frac{c \sqrt{\alpha}}{2 \sqrt{\pi}} \l k^2 \xi^2 / (2 + k^2 \xi^2) \r^{1/4}$. The data points in FIG.\ref{fig:setup} correspond to the estimated positions of the cross-overs between weak, strong and intermediate couplings, from plots as the one presented in FIG.\ref{fig:extendedRGpolaronMass}.

\section{Exact treatment of light impurities}
\label{sec:SmallImptMass}
In this section we develop a perturbation theory valid for small impurity masses $M \to 0$ and show that Lee-Low-Pines MF theory (see e.g. \cite{Lee1953,Devreese2013,BeiBing2009,Shashi2014RF}) becomes exact in this limit. We provide leading-order expressions for various quantities characterizing the small-mass polaron, which we derive in two spatial dimensions for concreteness (as relevant for photonic polarons \cite{Grusdt2015DSPP}). The generalization to arbitrary dimensions is straight forward.

To avoid approaching the subsonic- to supersonic transition, we keep the ratio $\beta := P/Mc$ of the conserved polaron momentum $P$ to the critical momentum $Mc$ of a non-interacting impurity fixed. The MF polaron is characterized by the coherent amplitude $\alpha_{\vec{k}}^\MF$ and the phonon momentum $P_{\rm ph}^\MF$, for which we find the following expansions,
\begin{equation}
\alpha_{\vec{k}}^{\MF} = - M \frac{V_k}{k^2/2} + \mathcal{O}(M^2), \qquad P_{\rm ph}^\MF = 4 M^3 \int d^2 \vec{k} ~ \frac{k_x^2 V_k^2}{k^4} \beta c + \mathcal{O}(M^4).
\label{eq:seriesExpansionsMF}
\end{equation}

The starting point for our analysis is the exact representation of the Fr\"ohlich Hamiltonian in Eq.\eqref{eq:HquantFlucDef}, based on an exact expansion around Lee-Low-Pines MF theory. We expand it in orders of $M$ now using Eq.\eqref{eq:seriesExpansionsMF} and find $\tilde{\mathcal{H}} = E_0^\MF + \sum_{n=-1}^\infty \tilde{\mathcal{H}}^{(n)}$, where $\tilde{\mathcal{H}}^{(n)} = \mathcal{O}(M^n)$. The leading order results are
\begin{flalign}
\tilde{\mathcal{H}}^{(-1)} &= \frac{1}{2 M} \l \int d^2 \vec{k} ~ \vec{k} \ad_{\vec{k}} \a_{\vec{k}} \r^2  \equiv \frac{\hat{\vec{Q}}^2}{2M}, \label{eq:HqTilde1} \\
\tilde{\mathcal{H}}^{(0)} &= \int d^2 \vec{k} ~ \l \omega_k - k_x \beta c \r \ad_{\vec{k}} \a_{\vec{k}} + \int d^2 \vec{k} d^2 \vec{k}' ~ \frac{\alpha_{\vec{k}}^\MF}{M} \vec{k} \cdot \vec{k}' \l \ad_{\vec{k}} \ad_{\vec{k}'} \a_{\vec{k}'} + \ad_{\vec{k}'} \a_{\vec{k}'} \a_{\vec{k}} \r , \label{eq:HqTilde2} \\
\tilde{\mathcal{H}}^{(1)} &= \frac{1}{2M} \int d^2 \vec{k} d^2 \vec{k}' ~ \alpha_{\vec{k}}^\MF \alpha_{\vec{k}'}^\MF \vec{k} \cdot \vec{k}' \l \ad_{\vec{k}} \ad_{\vec{k}'} + 2 \ad_{\vec{k}} \a_{\vec{k}'} + \a_{\vec{k}} \a_{\vec{k}'} \r. \label{eq:HqTilde3}
\end{flalign}

We can readily solve the leading order correction $\tilde{\mathcal{H}}^{(-1)}$, which is diagonal in the phonon Fock basis $\ket{n_{\vec{k}}}$. Because it depends only on the total phonon momentum $\hat{\vec{Q}}$, it produces highly degenerate sets of (excited) eigenstates. This degeneracy is broken by the first term in $\tilde{\mathcal{H}}^{(0)}$, which is still diagonal in the Fock basis. Therefore to order $M^0$ the polaron ground state corresponds to the phonon vacuum $\ket{0}$ (after expanding around the MF polaron). Thus the second term in Eq.\eqref{eq:HqTilde2} has no effect on the ground state, because it vanishes when acting on $\ket{0}$.

The leading order corrections to the ground state are derived from $\tilde{\mathcal{H}}^{(1)}$. For the ground state energy we find
\begin{equation}
\tilde{E} = E_0^\MF - \frac{1}{2 M} \int d^2 \vec{k} d^2 \vec{k}' ~ \frac{ \l \vec{k} \cdot \vec{k}'\r^2 |\alpha_{\vec{k}}^\MF|^2 |\alpha_{\vec{k}'}^\MF|^2 }{2 M \left[ \omega_k - k_x \beta c + \omega_{k'} - k_x' \beta c \right] + \l \vec{k} + \vec{k}' \r^2} =  E_0^\MF + \mathcal{O}(M^3).
\end{equation}
By expanding the ground state energy around $P=0$ we derive the leading order correction to the polaron mass,
\begin{equation}
\frac{1}{M_{\rm p}} = \frac{1}{M_{\rm p}^\MF} - 2^7 M^3 \int d^2 \vec{k} d^2 \vec{k}' ~ \frac{\l \vec{k} \cdot \vec{k}' \r^2 V_k^2 V_{k'}^2}{k^4 k^{' 4}} \left[ \frac{\l k_x + k_x' \r \l  \frac{k_x}{k^2} + \frac{k_x'}{k^{' 2}} \r }{\l \vec{k} + \vec{k}' \r^4 }  + 2 \frac{k_x k_x'}{k^2 k^{'2} \l \vec{k} + \vec{k}' \r^2 }  \right].
\end{equation}
They contribute to order $M^3$ only, whereas the MF polaron mass obeys $1/M_{\rm p}^\MF = 1/ M + \mathcal{O}(M)$. Thus MF theory also provides the correct leading order result for the polaron mass in the limit $M \to 0$.

\section{Extended all-coupling RG approach}
\label{sec:extRG}
In this section we formulate the extended RG in detail and derive the flow equations introduced in Sec.\ref{eq:overviewExtRG}. In parts the derivation is identical to the perturbative RG of Ref.\cite{Grusdt2015RG}. The starting point is the universal polaron Hamiltonian in $d$ spatial dimensions, see Eq.\eqref{eq:HquantFluc2}.

First we apply the same RG step as described in Ref. \cite{Grusdt2015RG}. Here we only give a brief summary and introduce our notations. The following infinitesimal unitary transformation,
\begin{equation}
\U_\Lambda =  \exp \l \int_\f d^d \vec{k} ~ \left[ \F_{\vec{k}}^\dagger \a_{\vec{k}} - \F_{\vec{k}} \ad_{\vec{k}} \right] \r,
\label{eq:defU}
\end{equation}
is used to decouple fast- from slow phonon degrees of freedom perturbatively in $\Omt_{\vec{k}}^{-1}$. As shown in Ref. \cite{Grusdt2015RG}, the choice $\F_{\vec{k}} =  \frac{ \tilde{\alpha}_{\vec{k}} }{\Omt_{\vec{k}}}  k_\mu \mathcal{M}_{\mu \nu}^{-1} \int_\s d^d \vec{p} ~  p_\nu  \Gt_{\vec{p}}  + \mathcal{O}(\Omt_{\vec{k}}^{-2})$ achieves this goal. It leads to the following transformation of the Hamiltonian,
\begin{equation}
\Ud_\Lambda \tilde{\mathcal{H}}_P \U_\Lambda  = \int_\f d^d \vec{k} ~ \ad_{\vec{k}} \a_{\vec{k}} \l \Omt_{\vec{k}} + \hat{\Omega}_\s(\vec{k}) \r + \H_\s +  \delta \H_\s,
\end{equation}
up to corrections of order $\mathcal{O}(\Omt_{\vec{k}})^{-2}$. Here we introduced $\hat{\Omega}_\s(\vec{k}) = k_\mu \mathcal{M}_{\mu \nu}^{-1} \int d^d \vec{p} ~ p_\nu \Gt_{\vec{p}}$. Most importantly, the slow phonon Hamiltonian is renormalized by
\begin{equation}
\delta \H_\s =  - \int_\f d^d \vec{k} ~ \frac{1}{\Omt_{\vec{k}}} \left[ \tilde{\alpha}_{\vec{k}}   k_\mu \mathcal{M}_{\mu \nu}^{-1} \int_s d^d \vec{p} ~ p_\nu \Gt_{\vec{p}} \right]^2 
+ \int_\f d^d \vec{k} ~ \frac{k_\mu  \mathcal{M}_{\mu \nu}^{-1} k_\nu }{2} \tilde{\alpha}_{\vec{k}}^2  + \mathcal{O}(\Omt_{\vec{k}})^{-2}.
\label{eq:renHs}
\end{equation}

Comparison to the universal Hamiltonian \eqref{eq:HquantFluc2} shows that the first term in Eq. \eqref{eq:renHs} gives rise to mass renormalization. The renormalized mass after the RG step reads
\begin{equation}
\tilde{\mathcal{M}}_{\mu \nu}^{-1} = \mathcal{M}_{\mu \nu}^{-1} - 2 \mathcal{M}_{\mu \lambda}^{-1}  \int_\f d^d \vec{k} ~ \frac{\tilde{\alpha}_{\vec{k}}^2 }{\Omt_{\vec{k}}} k_\lambda k_\sigma  ~\mathcal{M}_{\sigma \nu}^{-1},
\end{equation}
leading to the RG flow equation \eqref{eq:gsFlowM} for the tensorial mass. The second term in Eq.\eqref{eq:renHs} gives rise to an RG flow of the ground state energy, $\delta E_0^{(1)} = - \int_\s d^d \vec{p} ~ \frac{p_\mu  \mathcal{M}_{\mu \nu}^{-1} p_\nu }{2} \tilde{\alpha}_{\vec{p}}^2$. Here we made use of the relation $:\Gt_{\vec{k}} \Gt_{\vec{k}'}: = \Gt_{\vec{k}} \Gt_{\vec{k}'} - \delta \l \vec{k} - \vec{k}' \r \left[  \Gt_{\vec{k}} + |\tilde{\alpha}_{\vec{k}} |^2 \right]$ to write the initial Hamiltonian $\tilde{\mathcal{H}}_P(\Lambda)$ in a non-normal ordered form.

Using these identifications we can bring the renormalized Hamiltonian $\H_\s' = \H_\s + \delta \H_\s$ into the following form,
\begin{multline}
\H_\s'  = E_0 (\Lambda) + \delta E_0^{(1)} + \int_\s d^d \vec{p} ~ d^d \vec{p}' ~ \frac{1}{2}  p_\mu \tilde{\mathcal{M}}_{\mu \nu}^{-1} p_\nu'  ~ \Gt_{\vec{p}} \Gt_{\vec{p}'} + \int_\s d^d \vec{p} \left[ \ad_{\vec{p}} \a_{\vec{p}} \Omt_{\vec{p}} - \frac{1}{2} p_\mu \tilde{\mathcal{M}}_{\mu \nu}^{-1} p_\nu \Gt_{\vec{p}} \right] + \\
+ \int_\s d^d \vec{p}  \frac{1}{2} p_\mu \left[  \tilde{\mathcal{M}}_{\mu \nu}^{-1} -  \mathcal{M}_{\mu \nu}^{-1}  \right] p_\nu \Gt_{\vec{p}}.
\label{eq:HquantFlucRen}
\end{multline}
The first line is almost of the universal form \eqref{eq:HquantFluc2} again, except that the renormalization of $\Omt_{\vec{k}}$ is still missing and normal-ordering has not been performed. We will now show that it is provided by the terms in the second line, which will also lead to a renormalization of the MF amplitude $\tilde{\alpha}_{\vec{k}}$ entering the definition of $\Gt_{\vec{k}}$.

So far our analysis was completely equivalent to the RG procedure of Ref.\cite{Grusdt2015RG}. The crucial step now is to deal with the terms in the last line of Eq.\eqref{eq:HquantFlucRen} by applying a MF shift which treats fully self-consistently the coupling to \emph{all} slow-phonon modes. To carry out the MF shift, we write the renormalized Hamiltonian \eqref{eq:HquantFlucRen} in normal-ordered form as
\begin{equation}
\H_\s'  = E_0' (\Lambda) + \int_\s d^d \vec{p} ~ d^d \vec{p}' ~ \frac{1}{2}  p_\mu \tilde{\mathcal{M}}_{\mu \nu}^{-1} p_\nu'  ~ : \Gt_{\vec{p}} \Gt_{\vec{p}'}  : + \int_\s d^d \vec{p} \left[ \ad_{\vec{p}} \a_{\vec{p}} \Omt_{\vec{p}} + \delta \tilde{W}_{\vec{p}} \Gt_{\vec{p}} \right].
\label{eq:HquantFlucForMFshiftRG}
\end{equation}
We find for the energy $E_0' (\Lambda) = E_0(\Lambda) + \int_\s d^d \vec{p} ~ \delta \tilde{W}_{\vec{p}} \l \tilde{\alpha}_{\vec{p}} \r^2$, and we defined
\begin{equation}
\delta \tilde{W}_{\vec{p}} = \frac{1}{2} p_\mu \left[  \tilde{\mathcal{M}}_{\mu \nu}^{-1} -  \mathcal{M}_{\mu \nu}^{-1}  \right] p_\nu = \mathcal{O}(\delta \Lambda),
\end{equation}
which is related to the infinitesimal change of the renormalized mass. Therefore in what follows we may restrict ourselves to a perturbative treatment of such terms to first order in $\delta \Lambda$.
 
The RG-MF shift will be defined through
\begin{equation}
\hat{V}_\MF(\Lambda) = \exp \l \int_\s d^d \vec{p} ~  \delta \tilde{\alpha}_{\vec{p}} ~  \ad_{\vec{p}} -  \hc \r, \qquad \hat{V}_\MF^\dagger(\Lambda)  \a_{\vec{p}} \hat{V}_\MF(\Lambda) = \a_{\vec{p}} + \delta \tilde{\alpha}_{\vec{p}},
\label{eq:UMF-RG}
\end{equation}
where the renormalization of the MF amplitude $\delta \tilde{\alpha}_{\vec{p}} = \mathcal{O}(\delta \Lambda)$ is infinitesimal. This gives rise to renormalized $\G$-operators, $\Gtt_{\vec{p}} :=  \l \tilde{\alpha}_{\vec{p}} + \delta \tilde{\alpha}_{\vec{p}} \r  \l \a_{\vec{p}} + \ad_{\vec{p}} \r + \ad_{\vec{p}} \a_{\vec{p}}$, which are related to the initial operators by
\begin{equation}
\hat{V}_\MF^\dagger(\Lambda)  \Gt_{\vec{p}} \hat{V}_\MF(\Lambda) = \Gtt_{\vec{p}} + \l 2 \tilde{\alpha}_{\vec{p}} + \delta \tilde{\alpha}_{\vec{p}} \r \delta \tilde{\alpha}_{\vec{p}}.
\end{equation}

By canceling terms which are linear in $\a_{\vec{p}}$ in the resulting Hamiltonian $\hat{V}_\MF^\dagger \H_\s' \hat{V}_\MF$ (i.e. by minimizing the MF variational energy with respect to $\delta \tilde{\alpha}_{\vec{p}}$), we find
\begin{equation}
\delta \tilde{\alpha}_{\vec{p}} = - \frac{\tilde{\alpha}_{\vec{p}}}{\Omt_{\vec{p}}} \left[ \delta \tilde{W}_{\vec{p}} + 2 p_\mu \tilde{\mathcal{M}}^{-1}_{\mu \nu} \l \int_\s d^d \vec{p}' ~ p_\nu' \tilde{\alpha}_{\vec{p}'} \delta \tilde{\alpha}_{\vec{p}'}  \r \right].
\label{eq:RGMFequation}
\end{equation}
The resulting Hamiltonian now reads
\begin{equation}
\hat{V}_\MF^\dagger \H_\s' \hat{V}_\MF = E_0' (\Lambda) + \int_\s d^d \vec{p} ~ d^d \vec{p}' ~ \frac{1}{2}  p_\mu \tilde{\mathcal{M}}_{\mu \nu}^{-1} p_\nu'  ~ : \Gtt_{\vec{p}} \Gtt_{\vec{p}'}  : +  \int_\s d^d \vec{p}  ~ \ad_{\vec{p}} \a_{\vec{p}} ~ \Omt_{\vec{p}} \l 1 - \frac{\delta \tilde{\alpha}_{\vec{p}}}{\tilde{\alpha}_{\vec{p}}} \r.
\label{eq:MFRGrenHamiltonian1}
\end{equation}
From the last term we read off the renormalized frequency, $\tilde{\Omt}_{\vec{p}} = \Omt_{\vec{p}} \l 1 - \delta \tilde{\alpha}_{\vec{p}} / \tilde{\alpha}_{\vec{p}} \r$,
from which we can readily conclude that the renormalized MF amplitude after the complete RG step is
\begin{equation}
\tilde{\tilde{\alpha}}_{\vec{p}} := \tilde{\alpha}_{\vec{p}} + \delta \tilde{\alpha}_{\vec{p}} = - \frac{V_p}{\tilde{\Omt}_{\vec{p}}}.
\end{equation}

Notably all explicit energy corrections are of order $\delta \Lambda^2$ in Eq.\eqref{eq:MFRGrenHamiltonian1} and have thus been omitted in this equation. Nevertheless the RG-MF shift has an effect on the ground state energy, because the MF amplitude $\tilde{\alpha}_{\vec{p}}$ flows in the RG. Since Eq.\eqref{eq:MFRGrenHamiltonian1} is of the universal form \eqref{eq:HquantFluc2}, the corrections to the ground state energy are given by $\delta E_0 = E_0'(\Lambda) - E_0(\Lambda)$ in every RG step.

Finally we need to derive the RG flow of the renormalized dispersion $\Omt_{\vec{k}}$. To this end we first solve the RG-MF equation \eqref{eq:RGMFequation} by defining $\zeta_\mu := 2 \int_\s d^d \vec{p} ~ p_\mu \tilde{\alpha}_{\vec{p}} \delta \tilde{\alpha}_{\vec{p}}$. Plugging Eq.\eqref{eq:RGMFequation} back into this definition yields
\begin{equation}
\zeta_\mu = - 2 \int_\s d^d \vec{p} ~ p_\mu \frac{\tilde{\alpha}_{\vec{p}}^2}{\Omt_{\vec{p}}} \left[ \delta \tilde{W}_{\vec{p}} + p_\mu \mathcal{M}_{\mu \nu}^{-1} \zeta_\nu \right].
\end{equation}
This is a linear equation for $\zeta_\nu$ which we solve by assuming that the total polaron momentum $\vec{P} = P \vec{e}_x$ points along the $x$-direction. In this case by symmetry considerations $\zeta_\mu = \delta_{\mu x} \zeta_x$ and we obtain
\begin{equation}
\zeta_x = - \frac{\int_\s d^d \vec{p} ~ p_x  \frac{\tilde{\alpha}_{\vec{p}}^2}{\Omt_{\vec{p}}} p_\mu  \l \tilde{\mathcal{M}}_{\mu \nu}^{-1} - \mathcal{M}_{\mu \nu}^{-1} \r  p_\nu }{1 + 2 \mathcal{M}_{xx}^{-1} \int_\s d^d \vec{p} ~ p_x^2 \frac{\tilde{\alpha}_{\vec{p}}^2}{\Omt_{\vec{p}}}}.
\label{eq:zetaXres}
\end{equation}

Thus the renormalization of the dispersion is given by
\begin{equation}
\tilde{\Omt}_{\vec{p}} - \Omt_{\vec{p}} = \frac{1}{2} p_\mu \l \tilde{\mathcal{M}}_{\mu \nu}^{-1} - \mathcal{M}_{\mu \nu}^{-1} \r p_\nu + p_x \mathcal{M}_{xx}^{-1} \zeta_x.
\label{eq:deltaOmt}
\end{equation}
The first term on the right hand side describes the renormalization of the effective mass in the dispersion. The second term describes the RG flow of the coupling constant $\kappa_\nu$.

\section{Summary and Outlook}
\label{sec:SummaryOutlook}
In this article we have extended the wavefunction-based renormalization group theory of Fr\"ohlich polarons \cite{Grusdt2015RG} by including an RG flow of the MF solution. We demonstrated that the extended RG provides an efficient all-coupling theory of the polaron, describing correctly both limits of weak and strong couplings. For sufficiently light impurities we find an extended regime of intermediate couplings, where phonon correlations in the polaron cloud become important. This signals a break-down of Feynman's celebrated all-coupling  theory, which yields poorer variational energies and can not capture the intermediate coupling regime.  Using the extended RG developed in this article we calculated the full phase diagram of the Bogoliubov-Fr\"ohlich Hamiltonian. 

 In our method we combined the use of lowest-order perturbation theory and non-perturbative MF theory in every single RG step for deriving RG flow equations. We start from the MF solution, include quantum fluctuations and take into account the renormalization of the MF solution at lower energies. We expect that this approach is of much wider applicability, beyond the Fr\"ohlich Hamiltonian and even beyond polaron problems. It should be a useful technique to obtain not only universal properties of the effective low-energy theory, but also quantitative results for the coupling constants in the effective Hamiltonian. 

In the extended RG we obtain a flow of the effective Hamiltonian with the UV cut-off. After decoupling fast and slow phonons, the fast phonons are described by a simple quadratic Hamiltonian. In a forthcoming work we will show how this allows to extend our RG scheme further to solve non-equilibrium problems \cite{Grusdt2015dRG}.

\section*{Acknowledgements}
The author gratefully acknowledges Michael Fleischhauer for supporting this work. He is thankful for inspiring and useful discussions with Michael Fleischhauer, Eugene Demler, Yulia Shchadilova, Alexey Rubtsov, Wim Casteels and Richard Schmidt. Financial support from the Moore Foundation and from the DFG through SFB-TR 49 is gratefully acknowledged.\\

\def\bibsection{\section*{\refname}} 


%

\end{document}